 \documentclass[aps,showpacs,amsmath,amssymbd,dvips,showkeys,preprintnumbers]{revtex4}

\usepackage{graphicx}
\usepackage{color}

\def \be  {\begin{equation}}
\def \ee  {\end{equation}}
\def \ee  {\end{equation}}
\def \bea {\begin{eqnarray}}
\def \eea {\end{eqnarray}}

\def \Tr  {\bf{Tr}}

\newcommand{\nn}{\nonumber}

\begin{document}

\preprint{ECTP-2014-01}
\preprint{WLCAPP-2014-01}

\title{On SU(3) effective models and chiral phase-transition}

\author{Abdel Nasser ~TAWFIK\footnote{The author(s) declare(s) that there is no conflict
of interests regarding the publication of this article.}}
\email{a.tawfik@eng.mti.edu.eg}
\affiliation{Egyptian Center for Theoretical Physics (ECTP), Modern University for Technology and  Information (MTI), 11571 Cairo, Egypt}
\affiliation{World Laboratory for Cosmology And Particle Physics (WLCAPP), Cairo, Egypt}

\author{Niseem MAGDY} 
\affiliation{World Laboratory for Cosmology And Particle Physics (WLCAPP), 11571 Cairo, Egypt}
\affiliation{Brookhaven National Laboratory (BNL) - Department of Physics
P.O. Box 5000, Upton, NY 11973-5000, USA}

\begin{abstract}
The sensitivity of the Polyakov Nambu-Jona-Lasinio (PNJL) model  as an effective theory of quark dynamics to the chiral symmetry has been utilized in studying the QCD phase-diagram. Also, the Poyakov linear sigma-model (PLSM), in which information about the confining glue sector of the theory was included through the Polyakov-loop potential. The latter is to be extracted from pure Yang-Mills lattice simulations. Furthermore, from the quasi-particle model (QPM) as an excellent approach to reproduce the lattice QCD calculations, the gluonic sector of QPM is  integrated to LSM in order to reproduce recent lattice calculations. 
The hadron resonance gas (HRG) model gives an excellent description for the thermal and dense evolution of various thermodynamic quantities and characterizes  the conditions deriving the chemical freeze-out at finite densities. Therefore, the HRG model is used in calculating the thermal and dense dependence of quark-antiquark condensate. 
We review PLSM, QLSM, PNJL and HRG with respect to their descriptions for the chiral phase-transition. We analyse the chiral order-parameter  $M(T)$, the normalized net-strange condensate $\Delta_{q,s}(T)$ and the chiral phase-diagram and compare the results with the lattice QCD.  We conclude that PLSM works perfectly in reproducing $M(T)$ and $\Delta_{q,s}(T)$. The HRG model reproduces $\Delta_{q,s}(T)$, while PNJL and QLSM seem to fail. These features and differences are present in the QCD chiral phase-diagram. 
The PLSM chiral boundary is located in upper band of the lattice QCD calculations and agree well with the freeze-out results deduced from the high-energy experiments and the thermal models. Also, we find that the chiral temperature calculated from from HRG model is larger than that from the PLSM. This is also larger than the freeze-out temperatures calculated in the lattice QCD and deduced from experiments and thermal models. The corresponding temperature $T$ and chemical potential $\mu$ sets are very similar to that of PLSM. This might be explained, because the chiral $T$ and $\mu$  are calculated using different order parameters; in HRG vanishing quark-antiquark condensate but in PLSM crossing (equalling) chiral condensates and Polyakov loop potentials. The latter assumed that the two phase transitions; chiral and deconfinement, take place at the same temperature. The earlier deals with the chiral phase-transition independently on the deconfinement one. Although the results from the two model PNJL and QLSM keep the same behavior, the chiral temperatures deduced from both of them are higher than that of PLSM and HRG. This might be interpreted due the very heavy quark masses implemented in both models.

\end{abstract}


\maketitle
\tableofcontents
\makeatletter
\let\toc@pre\relax
\let\toc@post\relax
\makeatother 

\section{Introduction}
\label{Introduction}

At large momentum scale, quantum chromodynamics (QCD) predicts asymptotic freedom \cite{J-Gross,D-Politzer} or a remarkable weakening in the running strong coupling. Accordingly, phase transition takes place from hadrons in which quarks and gluons are conjectured to remain confined (at low temperature and density) to quark-gluon plasma (QGP) \cite{N. Cabibbo,Collins}, in which quarks and gluons become deconfined (at high temperature and density) \cite{Rischke:2003mt}. Furthermore, at low temperature, the QCD chiral symmetry is spontaneously broken; $SU(N_f )_L \times SU(N_f )_R \rightarrow SU(N_f )_V$. In this limit, the chiral condensate 
remains finite below the critical temperature ($T_c$). The broken chiral symmetry is restored at high temperatures. The finite quark masses explicitly break QCD chiral symmetry.

Nambu-Jona-Lasinio (NJL) model \cite{Nambu:1961tp} describes well the hadronic degrees of freedom. Polyakov Nambu-Jona-Lasinio (PNJL) model \cite{Fukushima:2003fw,Ratti:2005jh,Fukushima:2008wg} takes into consideration the quark dynamics \cite{Hatsuda:1994pi} and has been utilized to study the QCD phase-diagram \cite{Asakawa:1989bq,Fujii:2003bz}. Also, linear-$\sigma$ model (LSM) \cite{Gell Mann:1960} can be used in mapping out the QCD phase-diagram.

Many studies have been performed on LSM like O(4) LSM \cite{Gell Mann:1960} at vanishing temperature, O(4)  LSM at finite temperature \cite{Lenaghan:1999si, Petropoulos:1998gt} and $SU(N_f)_R  \times SU(N_f)_L$ LSM for $N_f=2$, $3$ and $4$ quark flavors \cite{l, Hu:1974qb, Schechter:1975ju, Geddes:1979nd}. In order to obtain reliable results,  extended LSM to PLSM ] can be utilized, in which information about the confining glue sector of the theory is included in form of Polyakov-loop potential. The latter can be extracted from pure Yang-Mills lattice simulations \cite{Polyakov:1978vu, Susskind:1979up, Svetitsky:1982gs,Svetitsky:1985ye}. Also, the Polyakov linear sigma-model (PLSM), and Polyakov quark meson model (PQM) \cite{Schaefer:2007pw,Kahara:2008yg,Schaefer:2008ax} deliver reliable results. Furthermore, the quasi-particle model (QPM) \cite{kmpf1,qp18b} was suggested to reproduce the lattice QCD calculations \cite{QPMqcd1,QPMqcd2}, in which two types of actions are implemented; the lattice QCD simulations utilizing the standard Wilson action and the ones with renormalization-improved action. 

In the present work, we integrate the gluonic sector of QPM to LSM \cite{Tawfik:2014bna} (QLSM) in order to reproduce the recent lattice QCD calculations \cite{Borsanyi:2013bia}. In QLSM \cite{Tawfik:2014bna}, the Polyakov contributions to the gluonic interactions and to the confinement-deconfinement phase transition are entirely excluded. Instead we just add the gluonic part of QPM. Therefore, the quark masses should be very heavy. We shall comment on this, later on. In section \ref{sec:Results}, we outline the QLSM results. They are similar to that of PNJL. This might be interpreted due the very heavy quark masses implemented in both models. Similar approach has been introduced in \cite{MAS2006}. The authors described in inclusion of gluonic Polyakov loop, which is assumed to generate a large gauge invariance and lead to a remarkable modification in hadron thermodynamics. A quite remarkable bridging between PNJL model quantum and local Polyakov loop and HRG model has been introduced \cite{AMS2014}. A large suppression of the thermal effects has been reported and it was concluded that the center symmetry breaking becomes exponentially small with increasing the massed of constituent quarks. In other words, the chiral symmetry restoration becomes exponentially small with increasing the pion mass.

The hadron resonance gas (HRG) model gives a good description for the thermal and dense evolution of various thermodynamic quantities in the hadronic matter \cite{Karsch:2003vd,Karsch:2003zq,Redlich:2004gp,Tawfik:2004sw,Tawfik:2004vv,Tawfik:2006yq,Tawfik:2010uh,Tawfik:2010pt,Tawfik:2012zz}. Also, it has been successfully utilized to characterizing  the conditions deriving the chemical freeze-out at finite densities~\cite{Tawfik:2012si}. In light of this,  the HRG model can be well used in calculating the thermal and dense dependence of quark-antiquark condensate \cite{Tawfik:2005qh}. The HRG grand canonical ensemble includes two important features \cite{Tawfik:2004sw}; the kinetic energies and the summation over all degrees of freedom and energies of  the resonances. On other hand, it is known that the formation of resonances can only be achieved through strong interactions~\cite{Hagedorn:1965st}; {\it resonances (fireballs) are composed of further resonances (fireballs), which in turn consist of  resonances (fireballs) and so on}. In other words, the contributions of the hadron resonances to the partition function are the same as that of free  (non-interacting) particles with  an effective mass. At temperatures comparable to the resonance half-width, the effective mass approaches the physical one \cite{Tawfik:2004sw}. Thus, at high temperatures, the strong interactions are conjectured to be taken into consideration through the inclusion of heavy resonances. It is found that the hadron resonances with masses up to $2\;$GeV are representing suitable constituents for the partition function ~\cite{Karsch:2003vd,Karsch:2003zq,Redlich:2004gp,Tawfik:2004sw,Tawfik:2004vv,Tawfik:2006yq,Tawfik:2010uh,Tawfik:2010pt,Tawfik:2012zz}. Such a way, the singularity expected at the Hagedorn temperature~\cite{Karsch:2003zq,Karsch:2003vd} can be avoided and the strong interactions are assumed to be taken into consideration. Nevertheless, validity of the HRG model is limited to  the temperatures below the critical one, $T_c$.

In the present paper, we review PLSM, QLSM, PNJL and HRG with respect to their descriptions for the chiral phase-transition.  We analyse the chiral order-parameter $M(T)$, the normalized net-strange condensate $\Delta_{q,s}(T)$ and the chiral phase-diagram and compare the results with the lattice QCD \cite{LQCD1,Schmidt:2010ss,Borsanyi:2010zi}. The present work is organized as follows. In section \ref{Model}, we introduce the different approaches SU(3) PLSM \cite{Tawfik:2014uka} (section \ref{PLSM}), QLSM (section \ref{LSM+QPM}), PNJL (section \ref{PNJL}) and HRG (section \ref{HRG:main}). The corresponding mean field approximations are also outlined. Section \ref{sec:Results} is devoted to the results. The conclusions and outlook shall be given in section \ref{sec:conclusion}.

\section{SU(3) effective models}
\label{Model}


\subsection{Polyakov Linear Sigma-Model (PLSM)}
\label{PLSM}

As discussed in Ref. \cite{Tawfik:2014uka,Tawfik:2014bna}. the Lagrangian of LSM with $N_f =3$  quark flavors and $N_c=3$ (for quarks, only) color degrees and with quarks coupled to Polyakov loop dynamics was introduced in \cite{Schaefer:2008ax,Mao:2010},
\begin{eqnarray}
\mathcal{L}=\mathcal{L}_{chiral}-\mathbf{\mathcal{U}}(\phi, \phi^*, T). \label{plsmI}
\end{eqnarray}
where the chiral part of the Lagrangian  of the SU(3)$_{l}\times$ SU(3)$_{R}$ symmetric linear sigma-model Lagrangian with $N_f =3$ is \cite{Lenaghan,Schaefer:2008hk} $\mathcal{L}_{chiral}=\mathcal{L}_q+\mathcal{L}_m$. The first  term is the fermionic part, Eq. (\ref{lfermion1}) with a flavor-blind Yukawa coupling $g$ of the quarks. The second term is the mesonic contribution, Eq. (\ref{lmeson1})   
\begin{eqnarray}
\mathcal{L}_q &=& \sum_f \overline{\psi}_f(i \gamma^{\mu} 
D_{\mu}-gT_a(\sigma_a+i \gamma_5 \pi_a))\psi_f, \label{lfermion1} \\
\mathcal{L}_m &=&
\mathrm{Tr}(\partial_{\mu}\Phi^{\dag}\partial^{\mu}\Phi-m^2
\Phi^{\dag} \Phi)-\lambda_1 [\mathrm{Tr}(\Phi^{\dag} \Phi)]^2 
-\lambda_2 \mathrm{Tr}(\Phi^{\dag}
\Phi)^2+c[\mathrm{Det}(\Phi)+\mathrm{Det}(\Phi^{\dag})]
+\mathrm{Tr}[H(\Phi+\Phi^{\dag})].  \label{lmeson1}
\end{eqnarray}
The summation $\sum_f$ runs over the three flavors (f=1, 2, 3 for the three quarks u, d, s). The flavor-blind Yukawa coupling $g$ should couple the quarks to the mesons. The coupling of the quarks to the Euclidean gauge field  $A_{\mu}=\delta_{\mu 0}A_0$ is given via the covariant derivative $D_{\mu}=\partial_{\mu}-i A_{\mu}$ \cite{Polyakov:1978vu,Susskind}. In Eq. (\ref{lmeson1}), $\Phi$ is a complex $3 \times 3$ matrix which depends on the $\sigma_a$ and $\pi_a$ \cite{Schaefer:2008hk}, where  $\gamma^{\mu}$ are  Dirac $\gamma$ matrices, $\sigma_a$ are the scalar mesons and  $\pi_a$ are the pseudoscalar mesons. 
\begin{eqnarray}
\Phi= T_a \phi _{a} =T_a(\sigma_a+i\pi_a),\label{Phi}
\end{eqnarray}
where $T_a=\lambda_a/2$ with $a=0, \cdots, 8$ are the nine generators of the U(3) symmetry group and $\lambda_a$ are the eight Gell-Mann matrices \cite{Gell Mann:1960}. The chiral symmetry is  explicitly broken through
 \begin{eqnarray}
H = T_a h_a.\label{H}
\end{eqnarray}
which is a $3 \times 3$ matrix with nine parameters $h_a$.  Three finite condensates $\bar{\sigma_0}$, $\bar{\sigma_3}$ and $\bar{\sigma_8}$ are possible, because the finite values of vacuum expectation of $\Phi$ and $\bar{\Phi}$ are conjectured to carry the vacuum quantum numbers and the diagonal components of the explicit symmetry breaking term, $h_a$, where $h_0 \neq 0$, $h_3=0$ and $h_8 \neq 0$, and squared tree level mass of the mesonic fields $m^2$, two possible coupling constants $\lambda_1$  and $\lambda_2$, Yukawa coupling $g$ and a cubic coupling constant $c$ can be estimated as follows. $c=4807.84~$MeV, $h_1=(120.73)^3~$MeV$^3$, $h_s=(336.41)^3~$MeV$^3$, $m^2=-(306.26)^2$MeV$^2$, $\lambda _1=13.48$ and $\lambda _3=46.48$ and  $g=6.5$.

In order to get a good  analysis it is more convenient to convert the condensates $\sigma_0$ and $\sigma_8$ into a pure non-strange $\sigma_x$ and strange $\sigma_y$ condensates \cite{Kovacs:2006}
\bea
\label{sigms}
\left( {\begin{array}{c}
\sigma _x \\
\sigma _y
\end{array}}
\right)=\frac{1}{\sqrt{3}} 
\left({\begin{array}{cc}
\sqrt{2} & 1 \\
1 & -\sqrt{2}
\end{array}}\right) 
\left({ \begin{array}{c}
\sigma _0 \\
\sigma _8
\end{array}}
\right).
\eea

The second term in Eq. (\ref{plsmI}) $\mathbf{\mathcal{U}}(\phi, \phi^*, T)$ represents Polyakov-loop effective potential \cite{Polyakov:1978vu}, which agrees well with the non-perturbative lattice QCD simulations and should have $Z(3)$ center symmetry as pure gauge QCD Lagrangian does \cite{Ratti:2005jh,Schaefer:2007pw}. In the present work, we use the potential $U(\phi, \phi^{*},T)$ as a polynomial expansion in $\phi $ and $ \phi^{*}$ \cite{Ratti:2005jh,Roessner:2007,Schaefer:2007d,Fukushima:2008wg}
 \begin{eqnarray}
\frac{\mathbf{\mathcal{U}}(\phi, \phi^*, T)}{T^4}=-\frac{b_2(T)}{2}|\phi|^2-\frac{b_3
}{6}(\phi^3+\phi^{*3})+\frac{b_4}{4}(|\phi|^2)^2, \label{UloopI}
\end{eqnarray}
where 
\begin{eqnarray}
b_2(T)=a_0+a_1\left(\frac{T_0}{T}\right)+a_2\left(\frac{T_0}{T}\right)^2+a_3\left(\frac{T_0}{T}\right)^3. 
\end{eqnarray}
 
In order to reproduce pure gauge lattice QCD thermodynamics and the behavior of the Polyakov loop as a function of 
temperature, we use the parameters $a_0=6. 75$,  $a_1=-1. 95$,  $a_2=2. 625$, $a_3=-7. 44$,  $b_3=0.75$ and $b_4=7.5$.  For a much better agreement with the lattice QCD results, the deconfinement temperature $T_0$  in pure gauge sector is fixed at $270$ MeV.

\subsubsection{Polyakov Linear Sigma-Model (PLSM) in Mean Field Approximation}
\label{PLSM:main}

In thermal equilibrium the grand partition function can be defined by using a path integral over the quark, anti-quark and meson fields 
\begin{eqnarray} \label{MFAEQ}
\mathcal{Z} &=& \mathrm{Tr \,exp}[-(\hat{\mathcal{H}}-\sum_{f=u, d, s}
\mu_f \hat{\mathcal{N}}_f)/T] 
= \int\prod_a \mathcal{D} \sigma_a \mathcal{D} \pi_a \int
\mathcal{D}\psi \mathcal{D} \bar{\psi} \mathrm{exp} \left[ \int_x
(\mathcal{L}+\sum_{f=u, d, s} \mu_f \bar{\psi}_f \gamma^0 \psi_f )
\right],
\end{eqnarray}
where $\int_x\equiv i \int^{1/T}_0 dt \int_V d^3x$ and $V$ is the volume of the system. $\mu_f$ is the chemical potential for $f=(u, d, s)$. We consider symmetric quark matter and define a uniform blind chemical potential $\mu_f \equiv \mu_{u, d}=\mu_s$. Then, we evaluate the partition function in the mean field approximation  \cite{Schaefer:2008hk,Scavenius:2000qd}.  We can use standard methods \cite{Kapusta:2006pm} in order to calculate the integration. This gives the effective potential for the mesons. 

We define the thermodynamic potential density of PLSM  as 
\begin{eqnarray}
\Omega(T, \mu)=\frac{-T \mathrm{ln}
\mathcal{Z}}{V}=U(\sigma_x, \sigma_y)+\mathbf{\mathcal{U}}(\phi, \phi^*, T)+\Omega_{\bar{\psi}
\psi}. \label{potentialI}
\end{eqnarray}
Assuming degenerate light quarks, i.e. $q\equiv u, d$, the quarks and anti-quarks contribution potential is given as \cite{Mao:2010}
\begin{eqnarray} \label{qqpotioI}
\Omega_{\bar{\psi} \psi} &=& -2 T N_q \int \frac{d^3\vec{p}}{(2
\pi)^3} \{ \mathrm{ln} [ 1+3(\phi+\phi^* e^{-(E_q-\mu)/T})\times
e^{-(E_q-\mu)/T}+e^{-3 (E_q-\mu)/T}] \nonumber \\&&  +\mathrm{ln} [
1+3(\phi^*+\phi e^{-(E_q+\mu)/T})\times e^{-(E_q+\mu)/T}+e^{-3
(E_q+\mu)/T}] \} \nonumber \\&& -2 T N_s \int \frac{d^3\vec{p}}{(2
\pi)^3} \{ \mathrm{ln} [ 1+3(\phi+\phi^* e^{-(E_s-\mu)/T})\times
e^{-(E_s-\mu)/T}+e^{-3 (E_s-\mu)/T}] \nonumber \\&&  +\mathrm{ln} [
1+3(\phi^*+\phi e^{-(E_s+\mu)/T})\times e^{-(E_s+\mu)/T}+e^{-3
(E_s+\mu)/T}] \},
\end{eqnarray}
where $N_q=2$, $N_s=1$, and the valence quark and antiquark energy for light and strange quark, $E_q=\sqrt{\vec{p}^2+m_q^2}$ and $E_s=\sqrt{\vec{p}^2+m_s^2}$, respectively. Also, as per \cite{Kovacs:2006} the light quark sector, Eq. (\ref{sqmass}),  decouples from the strange quark sector ($m_s$) and light quark mass $m_q$ gets simplified in this new basis to
\begin{eqnarray}
m_q &=& g \frac{\sigma_x}{2}, \label{qmass} \\
m_s &=& g \frac{\sigma_y}{\sqrt{2}}.  \label{sqmass}
\end{eqnarray} 

The purely mesonic potential is given as
 \begin{eqnarray}
U(\sigma_x, \sigma_y) &=& \frac{m^2}{2} (\sigma^2_x+\sigma^2_y)-h_x \sigma_x-h_y \sigma_y-\frac{c}{2\sqrt{2}} \sigma^2_x \sigma_y  + \frac{\lambda_1}{2} \sigma^2_x \sigma^2_y 
+ \frac{1}{8} (2 \lambda_1 +\lambda_2)\sigma^4_x + \frac{1}{4} (\lambda_1+\lambda_2)\sigma^4_y. \label{UpotioI}
\end{eqnarray}
We notice that the sum in Eqs. (\ref{UloopI}), (\ref{qqpotioI}) and (\ref{UpotioI}) give the thermodynamic potential density similar to Eq. (\ref{potentialI}), which has seven parameters $m^2, h_x, h_y, \lambda_1, \lambda_2, c$ and $ g$, two unknown condensates $\sigma_x$ and $ \sigma_y$ and two order parameters for the deconfinement $\phi$ and $\phi^*$. The six parameters $m^2, h_x, h_y, \lambda_1, \lambda_2 $ and $ c$  are fixed in the vacuum by six experimentally known quantities \cite{Schaefer:2008hk}. In order to evaluate the unknown parameters $\sigma_x$, $ \sigma_y$, $\phi$ and $\phi^*$, we minimize the thermodynamic potential, Eq. (\ref{potentialI}), with respective to $\sigma_x$, $ \sigma_y$, $\phi$ and $\phi^*$, respectively. Doing this, we obtain a set of four equations of motion
\begin{eqnarray}\label{cond1}
\frac{\partial \Omega}{\partial \sigma_x}= \frac{\partial
\Omega}{\partial \sigma_y}= \frac{\partial \Omega}{\partial
\phi}= \frac{\partial \Omega}{\partial \phi^*}\mid_{min} =0,
\end{eqnarray}
where {\it min} means $\sigma_x=\bar{\sigma_x}, \sigma_y=\bar{\sigma_y}, \phi=\bar{\phi}$ and $\phi^*=\bar{\phi^*}$ are the global minimum. 

\begin{eqnarray}
\Omega_{PLSM} &=& {\cal {U}}(\phi,\bar \phi,T)+\frac{m^2}{2} (\sigma^2_x+\sigma^2_y)-h_x
\sigma_x-h_y \sigma_y-\frac{c}{2\sqrt{2}} \sigma^2_x \sigma_y  + \frac{\lambda_1}{2} \sigma^2_x \sigma^2_y  
+ \frac{1}{8} (2 \lambda_1 +\lambda_2)\sigma^4_x + \frac{1}{4} (\lambda_1+\lambda_2)\sigma^4_y   \nn \\
&-&
2 T N_q \int \frac{d^3\vec{p}}{(2 \pi)^3} \{ \mathrm{ln} [ 1+3(\phi+\phi^* e^{-(E_q-\mu)/T})\times
e^{-(E_q-\mu)/T}+e^{-3 (E_q-\mu)/T}] \nonumber \\
&& + \mathrm{ln} [1+3(\phi^*+\phi e^{-(E_q+\mu)/T})\times e^{-(E_q+\mu)/T}+e^{-3 (E_q+\mu)/T}] \} \nonumber \\
&-& 2 T N_s \int \frac{d^3\vec{p}}{(2 \pi)^3} \{ \mathrm{ln} [ 1+3(\phi+\phi^* e^{-(E_s-\mu)/T})\times
e^{-(E_s-\mu)/T}+e^{-3 (E_s-\mu)/T}] \nonumber \\
&& + \mathrm{ln} [1+3(\phi^*+\phi e^{-(E_s+\mu)/T})\times e^{-(E_s+\mu)/T}+e^{-3 (E_s+\mu)/T}] \},  \label{plsm:omega}
\end{eqnarray}
Accordingly, the chiral order parameter  can be deduced as 
\begin{eqnarray}
M_{PLSM} &=& m_s \dfrac{\langle \bar{\psi}\psi \rangle_{PLSM}}{T^{4}} = \frac{m_s}{T^{4}} \dfrac{\partial  \Omega_{PLSM}}{\partial m_l}.
\label{Eq:M}
\end{eqnarray}
 

\subsection{Linear Sigma-Model and Quasi-Particle Sector (QLSM) }
\label{LSM+QPM}

When the Polyakov contributions to the gluonic interactions and to the confinement-deconfinement phase transition are entirely excluded, the Lagrangian of LSM with $N_f=3$  quark flavors and $N_c=3$ (for quarks, only) color degrees of freedom, where the quarks couple to the Polyakov loop dynamics, has been introduced in Ref. \cite{Schaefer:2008ax,Mao:2010}, 
\begin{eqnarray}
\mathcal{L}=\mathcal{L}_{chiral}-\mathbf{\mathcal{U}}(\phi, \phi^*, T). \label{plsm}
\end{eqnarray}
The main original proposal of the present work is the modification of Eq. (\ref{plsm})
\begin{eqnarray}
\mathcal{L}=\mathcal{L}_{chiral}-\mathbf{\mathcal{U}_g}(T,\mu), \label{newplsm}
\end{eqnarray}
where the chiral part of the Lagrangian $\mathcal{L}_{chiral}=\mathcal{L}_q+\mathcal{L}_m$ is of SU(3)$_{L}\times$ SU(3)$_{R}$ symmetry  \cite{Lenaghan,Schaefer:2008hk}. Instead of $\mathbf{\mathcal{U}}(\phi, \phi^*, T)$, the gluonic potential $\mathbf{\mathcal{U}_g}(T,\mu)$, which is similar to the gluonic sector of the quasi-particle model, is inserted, (review Eq. (\ref{Eq:Ug})). The Lagrangian with $N_f =3$ consists of two parts; fermionic and mesonic contributions, Eqs. (\ref{lfermion1}) and (\ref{lmeson1}), respectively.   

Some details about the quasi-particle model are in order. The model gives a good phenomenological description for lattice QCD simulation and treats the interacting massless quarks and gluons as non-interacting massive quasi-particles \cite{qusim1}. The corresponding degrees of freedom are treated in a similar way as the electrons in condensed matter theory \cite{paul}, i.e. the interaction with the medium provides the quasi-particles with dynamical masses. Consequently, most of the interactions can be taken into account. When confronting it to the lattice QCD calculations, the free parameters can be fixed.  The pressure at finite $T$ and $\mu$ is given as
\bea
p &=& \sum_{i=q,g}\, p_i - B(T,\mu), \qquad\qquad
p_i = \frac{g_i}{6\, \pi^2} \int_0^{\infty} k^4\, dk\, \frac{1}{E_i(k)} \left[f_i^+(k)+f_i^-(k)\right],  \label{Eq:pg}
\eea
where the function $B(T,\mu)$ stands for bag pressure at finite $T$ and $\mu$ which can be determined by thermodynamic self-consistency and $\partial p/\partial \Pi_a=0$; the stability of $p$ with respect to the self-energies ($\Pi_a$) and the distribution function for bosons and fermions, $\pm$ respectively is given as
\bea \label{Eq:fi}
f_i^{\pm}(k) &=& \frac{1}{\exp\left(\frac{E_i(k)\mp \mu}{T}\right)\pm 1}.
\eea
The quasi-particle dispersion relation can be approximated by the asymptotic mass shell expression near the light cone \cite{kmpf1,qp18b},
\bea \label{Eq:Ei}
E_i^2(k) &=& k^2 + m_i^2(T,\mu)=k^2 + \Pi_i(k; T,\mu) + (x_i\, T)^2, 
\eea
where $\Pi_i(k; T,\mu)$ is the self-energy at finite $T$ and $\mu$ and $x_i^2$ is a factor taking into account the mass scaling as used in the lattice QCD simulations. In other words, $x_i^2$ was useful when the lattice QCD simulations have been performed with quark masses heaver than the physical ones. In the present work, the gluon self-energies $\Pi_g(k; T,\mu)$ are relevant \cite{Bluhm}
\bea \label{pi:Ei}
\Pi_g(k; T,\mu)=\left(\left[3+\dfrac{N_f}{2}\right] T^2 + \dfrac{3}{2 \pi ^2} \sum_f \mu_{f}^2\right) \dfrac{G^2}{6},
\eea
where the effective coupling $G$ at vanishing chemical potential is given as,
\bea  \label{G:Ei}
G^2(T) &=& \left\{ \begin{array}{ll}
G^2_{\text{2loop}}(T), & T\geq T_c, \\
& \\
G^2_{\text{2loop}}(T)+b\left(1-\frac{T}{T_c}\right), & T< T_c.\end{array}
\right.,
\eea
And the two-loop effective coupling $ G^2_{\text{2loop}}(T) $ reads \cite{kmpf1}
\bea  
G^2_{\text{2loop}}(T) &=& \frac{16\, \pi^2}{\beta_0 \ln(\xi^2)} \left[1-2 \frac{\beta_1}{\beta_0^2} \frac{\ln(\ln(\xi^2))}{\ln(\xi^2)}\right], \qquad \xi = \lambda \frac{T-T_s}{T_c}, \label{Gloop:Ei}
\eea
and $T_s$ is a regulator at $T_c$. For The parameter $\lambda$ is used to adjust the scale as found in lattice QCD simulations. These two parameters are not very crucial in the present calculations. The regulator and scale are controlled by the condensates ($\sigma_x$ and $\sigma_y$) and the order parameters ($\phi$ and $\phi^*$), which are given as function of temperature and baryon chemical potential. The $\beta$ function \cite{betaf} depends on the QCD coupling  $G$, $\beta=\partial\, G/(\partial\, \ln(\Delta_{\mu}))$, with $\Delta_{\mu}$ is the energy scale.  It is obvious that the QCD coupling in Eqs. (\ref{G:Ei}) and (\ref{Gloop:Ei}) is given in $T$-dependence, only. In calculating $\beta=\partial\, G/(\partial\, \ln(\Delta_{\mu}))$ at finite $\mu$ it is apparently needed to extend $G$ to be $\mu$-dependent, as well. The two-loop perturbation estimation for $\beta$ functions gives
\bea
\beta_0 &=& \frac{1}{3} \left(11\, n_c - 2\, n_f\right), \\
\beta_1 &=& \frac{1}{6} \left(34\, n_c^2 - 13\, n_f\, n_c + 3\, \frac{n_f}{n_c}\right).
\eea

\subsubsection{Linear Sigma-Model and Quasi-Particle Sector (QLSM) in Mean Field Approximation}
\label{LSM+QPM:main}

As in section \ref{PLSM:main} and Eq. (\ref{MFAEQ}), we derive the thermodynamic potential density in the mean field approximation. This consists of three parts:  mesonic and quasi-gluonic potentials in additional to the quark potential 
\begin{eqnarray}
\Omega(T, \mu)=\frac{-T\, \ln (Z)}{V}=U(\sigma_x,\, \sigma_y) + U_{g}(T,\, \mu) + \Omega_{\bar{\psi}
\psi}. \label{potential}
\end{eqnarray}
\begin{itemize}
\item First, the quark potential part \cite{Schaefer:2008hk}
\begin{equation}
\label{eq:quark_pot}
\Omega_{\bar{q}q}(T,\mu_f) = d_q\, T \sum_{f=u,d,s}
\int\limits_0^\infty \! \frac{d^3 k}{(2\pi)^3} { \ln \left[1- n_{q,f}(T,\mu_f)\right] + \ln \left[1-n_{\bar{q},f}(T,\mu_f)\right] }
\end{equation}
It is obvious that  $\Omega_{\bar{\psi}\psi}$ is equivalent to $\Omega_{\bar{q}q}$.
The occupation quark/aniquarks numbers read, 
\begin{equation}
n_{q|\bar{q},f}\left(T,\mu_{f}\right)=\frac{1}{1+\exp\left[(E_{q,f}\pm \mu_f)/T\right]},
\end{equation} 
and antiquarks $n_{\bar q,f}(T,\mu_{f}) \equiv n_{q,f} (T,-\mu_{f})$, respectively. The number of internal quark degrees of freedom is denoted by $d_q=2$ and $N_{c}=6$ (for quarks and antiquarks). The energies are given as
\begin{equation}
E_{q,f}= \sqrt{k^2 + m_f^2},
\end{equation}
with the quark masses $m_f$ which is related to $m_q$ and $m_s$ for $u$-, and $d$- and $s$-quark, respectively. 
As given, the latter are proportional to the $\sigma$-fields
\begin{eqnarray}
m_q &=& g \frac{\sigma_x}{2}, \label{qmass2} \\
m_s &=& g \frac{\sigma_y}{\sqrt{2}} \label{sqmass2},
\end{eqnarray} 
where the Yukawa coupling $g=8.3$. The symbols for the chiral condensates, $\sigma_x$ and $\sigma_y$ for light- and strange-quarks, respectively, are kept as in literature.

\item Second, the purely mesonic potential part reads
\begin{eqnarray}
U(\sigma_x, \sigma_y) &=& \frac{m^2}{2} \left(\sigma^2_x+\sigma^2_y\right)-h_x
\sigma_x-h_y \sigma_y-\frac{c}{2\sqrt{2}} \sigma^2_x \sigma_y
+\frac{\lambda_1}{2} \sigma^2_x \sigma^2_y  
+ \frac{\left(2 \lambda_1 + \lambda_2\right)\sigma^4_x}{8} + \frac{\left(\lambda_1+\lambda_2\right)\sigma^4_y}{4}. \label{Upotio} \hspace*{8mm}
\end{eqnarray}

\item Third, the quasi-gluonic potential part is constructed from Eqs. (\ref{Eq:Ei}), (\ref{Eq:fi}) and (\ref{Eq:pg})
\bea 
U_g &=& -\frac{d_g}{6\, \pi^2} \int_0^{\infty} k^4\, dk \frac{1}{E_i}  \left[\frac{1}{\exp\left(\frac{E_i - \mu}{T}\right)- 1} + 
       \frac{1}{\exp\left(\frac{E_i + \mu}{T}\right)- 1}\right]. \label{Eq:Ug}
\eea 
In Eq. (\ref{Eq:Ug}), the degeneracy factor $ d_g = 8$ and two parameters $\lambda$ and $T_s$, which were given in Eq. (\ref{Gloop:Ei}), should be fixed in order to reproduce the lattice  QCD calculations. Here, we find that $\lambda=2.0 $ and $T_s=0.0~$MeV give excellent results. 

When adding the three potentials given in Eqs. (\ref{Eq:Ug}), (\ref{Upotio}) and (\ref{eq:quark_pot}), the thermodynamics and chiral phase translation can be analysed. The  resulting potential $\Omega_{QLSM}$ can be used to determine the normalized net-strange condensate  and chiral order parameter, Eq. (\ref{Eq:M}).
\begin{eqnarray} 
\Omega_{QLSM}&=&\frac{m^2}{2} (\sigma^2_x+\sigma^2_y) - h_x\, \sigma_x-h_y \sigma_y-\frac{c}{2\sqrt{2}} \sigma^2_x \sigma_y  + \frac{\lambda_1}{2} \sigma^2_x \sigma^2_y + \frac{1}{8} (2 \lambda_1 + \lambda_2)\sigma^4_x + \frac{1}{4} (\lambda_1+\lambda_2)\sigma^4_y \nn \\ 
&-& T \frac{d_q}{2 \pi^2} \int_0^{\infty}k^2\, dk \left[2\ln (1-f_q^-(T,\mu ))+2\ln (1-f_q^+(T,\mu )) 
+ \ln (1-f_s^{-}(T,\mu ))+\ln (1-f_s^{+}(T,\mu ))\right] \nn \\ 
&-& 3\, \pi ^2\, d_g \int_0^{\infty}k^4\, dk \left[ \left(e^{\frac{\omega _g(T,\mu )}{T}}-1\right) \omega _g(T,\mu )\right]^{-1}, \label{Pq}
\end{eqnarray}
where, 
\begin{eqnarray} 
f_q^{\pm} (T,\mu) &=& \frac{1}{e^{\frac{\sqrt{\frac{1}{4} g^2 \sigma _x(T,\mu ){}^2+k^2}\pm \mu }{T}}+1}, \label{var1}\\
f_s^{\pm} (T,\mu) &=& \frac{1}{e^{\frac{\sqrt{\frac{1}{2} g^2 \sigma _y(T,\mu ){}^2+k^2}\pm \mu }{T}}+1}, \label{var2}\\ 
\omega _g(T,\mu ) &=& \left[k^2+\frac{8\, \pi ^2 \left(\frac{9}{2\, \pi ^2}\, \mu+\left(\frac{N_f}{2}+3\right) T^2\right) \left(1-\frac{3 \left(34\, N_c^2-13\, N_c\, N_f+3\, \frac{N_f}{N_c}\right) \ln \left(\ln ^2\left(\xi^2\right)\right)}{(11\, N_c-2\, N_f)^2 \ln^2\left(\xi^2\right)}\right)}{(11\, N_c-2\, N_f) \ln ^2\left(\xi^2\right)}\right]^{1/2},  \label{var3}
\end{eqnarray} 
$\xi$ is function of $T$ and the quark masses should be very heavy. The QLSM results are similar to that of PNJL, section \ref{sec:Results}. This might be interpreted due the very heavy quark masses implemented in both models.

\subsection{Polyakov Nambu-Jona-Lassinio (PNJL) Model}
\label{PNJL}

The Lagrangian of PNJL reads \cite{OHP2006,Abhijit} 
\begin {align}
   {\cal L} &= {\sum_{f=u,d,s}}{\bar\psi_f}\gamma_\mu i D^\mu
             {\psi_f}-\sum_f m_{f}{\bar\psi_f}{\psi_f}
              +\sum_f \mu \gamma_0{\bar \psi_f}{\psi_f}\nonumber\\
       &+{\frac {g_S} {2}} \sum_{a=0,\ldots,8} [({\bar\psi} \lambda^a {\psi})^2+
            ({\bar\psi} i\gamma_5\lambda^a {\psi})^2]
       -{g_D} [det{\bar\psi_f}{P_L}{\psi_{f^\prime}}+det{\bar\psi_f}
            {P_R}{\psi_{f^\prime}}]\nonumber\\ 
        &-{\cal {U}}(\phi[A],\bar \phi[A],T)
\label{lag}
\end {align}
where the matrices $P_{L,R}=(1\pm \gamma_5)/2$ are chiral projectors, ${\cal {U}}(\phi[A],\bar \phi[A],T)$ is the Polyakov loop potential (Landau-Ginzburg type potential) and $D^\mu=\partial^\mu-i{A_4}\delta_{\mu 4}$ stand for gauge field interactions. The mass of a particular flavor is denoted by $m_f$, where $f=u,d,s$. The two coupling constants $g_D$ and $g_S$, $\lambda^a$ are Gell-Mann matrices \cite{Gell Mann:1960} and $\gamma_\mu$  are Dirac $\gamma$ matrices. The model is not renormalizable so that we have to use three-momentum cut-off regulator $\Lambda$ in order to keep quark loops finite. 

The Polyakov loop potential is given by \cite{ratti},
\begin{equation}
\frac  {{\cal U}(\phi, \bar \phi, T)}{  {T^4}}=-\frac {{b_2}(T)}{ 2}
                 {\bar \phi}\phi-\frac {b_3}{ 6}(\phi^3 + \bar \phi^3)
                 +\frac {b_4}{  4}{(\bar\phi \phi)}^2
\end{equation}
with
\begin {eqnarray}
\phi &=& \frac{Tr_c L}{N_c}, \qquad
{\bar \phi} = \frac{Tr_c L^\dagger}{N_c}, \qquad
{b_2}(T) = a_0+{a_1}\left(\frac { {T_0}}{ T}\right)+{a_2}\left(\frac {{T_0}}{ T}\right)^2+
              {a_3}\left(\frac {{T_0}}{T}\right)^3, \nn
\end {eqnarray}
$b_3$ and $b_4$ being constants and we choose the following fitting values for the potential parameters, 
$a_0=6. 75$, $a_1=-1. 95$,  $a_2=2. 625$, $a_3=-7. 44$,  $b_3 = 0.75$ $b_4=7.5$ and  $T_0=187$~MeV. These are adjusted to the pure gauge lattice data such that the equation of state and the Polyakov-loop expectation values are reproduced.

\subsubsection{Polyakov Nambu-Jona-Lassinio (PNJL) Model in Mean Field Approximation}
\label{NJL:main}

The thermodynamic potential density of PNJL is defined as
\begin {align} \label{PNJLpotuntial}
 \Omega &= {\cal {U}}[\phi,\bar \phi,T]+2{g_S}{\sum_{f=u,d,s}}
            {\sigma_f^2}-\frac {{g_D} }{2}{\sigma_u}
          {\sigma_d}{\sigma_s}-6{\sum_f}{\int_{0}^{\Lambda}}
     \frac {{d^3p}}{{(2\pi)}^3} E_{pf}\Theta {(\Lambda-{ |\vec p|})}\nonumber \\
       &-2{\sum_f}T{\int_0^\infty}\frac {{d^3p}}{{(2\pi)}^3}
       \ln\left[1+3\left(\phi+{\bar \phi}e^{\frac {-(E_{pf}-\mu)}{ T}}\right)
       e^{\frac {-(E_{pf}-\mu)}{ T}}+e^{\frac {-3(E_{pf}-\mu)}{ T}}\right]
\nonumber\\
       &-2{\sum_f}T{\int_0^\infty}\frac {{d^3p}}{{(2\pi)}^3}
        \ln\left[1+3\left({\bar \phi}+{ \phi}e^{\frac {-(E_{pf}+\mu)}{ T}}\right)
       e^{\frac {-(E_{pf}+\mu)}{ T}}+e^{ \frac {-3(E_{pf}+\mu)}{ T}}\right],
\end {align}
where $E_{pf}=\sqrt {p^2+M^2_f}$ is the single quasi-particle energy,
$\sigma_f^2=\sigma_u^2+\sigma_d^2+\sigma_s^2$ and from isospin symmetry, $\sigma_q=\sigma_u=\sigma_d$. In the above integrals, the vacuum integral has a cutoff $\Lambda$ 
whereas the medium dependent integrals have been extended to infinity. By the self-consistent gap equation,  the quark mass can be estimated,
\begin {equation}
  M_f =m_f - 2 g_S\, \sigma_f + \frac {{g_D}}{2}\, \sigma_{f+1}\; \sigma_{f+2},
\end {equation}
where $\sigma_f=\langle{\bar \psi_f} \psi_f\rangle$ denotes the chiral condensate of quark with flavor $f$ and other parameters are listed out in Tab. \ref{table2PNL} \cite{OHP2006,Abhijit}. For isospin symmetry,  we define the light and strange-quark masses as
\begin {eqnarray}
  M_s &=& m_s - 2 g_S\, \sigma_s + \frac {{g_D}}{2}\; \sigma_{q}^2, \\
  M_q &=& m_q - 2 g_S\, \sigma_q + \frac {{g_D}}{2}\; \sigma_{q}\;  \sigma_{s}.
\end {eqnarray}
Here, we notice the strong dependence on the $\sigma$-fields.

\begin{table}
\begin{center}
\begin{tabular}{|c|c|c|c|c|}
\hline
$ m_u$ [MeV] & $m_s$ [MeV]&$ \Lambda $ [MeV] & $g_S \Lambda^2 $&$ g_D \Lambda^5 $\\
\hline
$ 5.5 $&$ 134.758 $&$ 631.357 $&$ 3.664 $&$ 74.636$ \\
\hline
\end{tabular}
\caption{Parameters of the SU(3) PNJL model. \label{table2PNL}}
\end{center}
\end{table}

Now, we have all the PNJL Model parameters except $\sigma_{q}, \sigma_{s}, \phi$ and $\bar \phi$, which can be estimated from minimizing the thermodynamic potential, Eq. (\ref{PNJLpotuntial}), with respective to $ \sigma_{q}, \sigma_{s}, \phi$ and $\bar \phi$, respectively. Doing this, we obtain a set of four equations of motion
\begin{eqnarray}\label{cond1}
\frac{\partial \Omega}{\partial \sigma_x}= \frac{\partial \Omega}{\partial \sigma_y}= \frac{\partial \Omega}{\partial \phi}= \frac{\partial \Omega}{\partial {\bar \phi}}\mid_{min} =0.
\end{eqnarray}

Then, the potential of the PNJL model reads
\begin {align} 
 \Omega _{PNJL}  &= {\cal {U}}[\phi,\bar \phi,T]+2{g_S}{\sum_{f=u,d,s}}
            {\sigma_f^2}-\frac {{g_D} }{2}{\sigma_u}
          {\sigma_d}{\sigma_s}-6{\sum_f}{\int_{0}^{\Lambda}}
     \frac {{d^3p}}{{(2\pi)}^3} E_{pf}\Theta {(\Lambda-{ |\vec p|})}\nonumber \\
       &-2{\sum_f}T{\int_0^\infty}\frac {{d^3p}}{{(2\pi)}^3}
       \ln\left[1+3\left(\phi+{\bar \phi}e^{\frac {-(E_{pf}-\mu)}{T}}\right)
       e^{\frac {-(E_{pf}-\mu)}{T}}+e^{\frac {-3(E_{pf}-\mu)}{T}}\right]
\nonumber\\
       &-2{\sum_f}T{\int_0^\infty}\frac {{d^3p}}{{(2\pi)}^3}
        \ln\left[1+3\left({\bar \phi}+{ \phi}e^{\frac {-(E_{pf}+\mu)}{T}}\right)
       e^{\frac {-(E_{pf}+\mu)}{T}}+e^{ \frac {-3(E_{pf}+\mu)}{T}}\right]. \label{PNJLpotuntial}
\end {align}

Having completed the introduction of both PLSM and PNJL, it is in order now to discuss central $Z(3)$ symmetry related to the Polyakov loop. It has been shown that the SU(3) color-singlet has $Z(3)$ symmetry through the normalized character in the fundamental representation of SU(3), $\Phi(\theta_1,\theta_2)$. This becomes equivalent to an ensemble of Polyakov loop \cite{IAMRG}. Furthermore, it was concluded that $\Phi(\theta_1,\theta_2)$ can taken as an order parameter for color-confinement to color-deconfinement phase transition, i.e. the center symmetry is spontaneously broken at high temperatures.

Ref \cite{MHOB:2012} introduced an attempt to resolve some incongruities within NJL and PNJL. It was argued that by integrating corresponding extremum conditions, the thermodynamic potential is directly obtained, where the integration constant
can be fixed from Stefan-Boltzmann law. Keeping the regulator finite at finite temperature and chemical potential is the main advantage of this approach.

\subsection{Hadron Resonance Gas (HRG) Model}
\label{HRG:main}

Treating hadron resonances as a free (non-interacting) gas~\cite{Karsch:2003vd,Karsch:2003zq,Redlich:2004gp,Tawfik:2004sw,Tawfik:2004vv} is conjectured to give an accurate estimation for the thermodynamic pressure below $T_c$. It has been shown that thermodynamics of strongly interacting  system can also be approximated as an ideal gas composed of hadron resonances with masses $\le 2~$GeV ~\cite{Tawfik:2004sw,Vunog}, i.e. confined QCD matter (hadrons) is well modelled as a non-interacting gas of resonances. The grand canonical partition function reads
\bea
Z(T, \mu, V) &=& \Tr \left[ \exp^{\frac{\mu\, N-H}{T}}\right],
\eea
where $T$ ($\mu$) is temperature (chemical potential). The Hamiltonian ($H$) is given as the kinetic energies of the relativistic Fermi and Bose particles. 

The main motivation of using $H$ is that 
\begin{itemize}
\item it contains all relevant degrees of freedom of confined, {\it strongly interacting} QCD matter,
\item it {\it implicitly} includes interactions, especially the ones leading to formation of resonances and 
\item it gives a quite satisfactory description of the particle production in heavy-ion collisions. 
\end{itemize}
With these assumptions, the thermodynamics is resulted from {\it single-particle partition} functions $Z_i^1$ 
\bea 
\ln Z(T, \mu_i ,V) &=& \sum_i\pm \frac{V g_i}{2\pi^2}\int_0^{\infty} k^2\, \ln\left\{1 \pm \exp\left[\frac{\mu_i -\varepsilon_i(k)}{T}\right]\right\}\, d k, \label{eq:lnz1}
\eea 
where $\varepsilon_i(k)=(k^2+ m_i^2)^{1/2}$ is the $i-$th particle dispersion relation, $g_i$ is
spin-isospin degeneracy factor and $\pm$ stands for bosons and fermions, respectively.

For hadron resonances which not yet measured, experimentally, a parametrization for a total spectral weight has been proposed \cite{brnt} as a recent estimation for Hagedorn mass spectrum \cite{hgdrn}. In the present work, we merely include known (measured) hadron resonances with mass $\leq 2~$GeV. This mass cut-off is assumed to define the validity of HRG in modelling the hadronic phase. Resonances with heavier masses diverge all thermodynamic quantities at the Hagedorn temperature~\cite{Karsch:2003zq, Karsch:2003vd}. 

Very recently, it has been shown that indeed the viral expansion is a reliable way to include hadron resonances, because the phase shift is a directly-accessible quantity in experiments \cite{BGB2015}. For instance, for accurate isospin-averaged observables, the scalar-isoscalar $f_0(500)$ ($\sigma$ meson) resonance and scalar $K^*(800)$ should be not be included in the HRG model.

The HRG model has been used in calculating the higher-order moments of the particle multiplicity, in which a grand canonical partition function of an ideal gas with experimentally-observed states up to a certain mass cut-off is utilized \cite{Tawfik:2012si}. The HRG model has been successfully utilized in characterizing two different conditions generating the chemical freeze-out at finite densities, namely constant normalized-entropy density $s/T^3=7$ \cite{sT3p1,sT3p2,Tawfik:2005qn,Tawfik:2004ss}, constant product of kurtosis and variance $\kappa\, \sigma^2=0$ \cite{Tawfik:2013dba} and constant trace-anomaly $(\epsilon-3 p)/T^4=7/2$ \cite{Tawfik:2013eua}.  As introduced in  \cite{Tawfik:2004ss}, the third freeze-out conditions, which is characterized by constant $s/T^3$ is accompanied by constant $s/n$.  

Our HRG model was used to study the possible differences between the behavior of light 
$\langle\bar{q}q\rangle=\langle\bar{u}u\rangle=\langle\bar{d}d\rangle$ and strange $\langle\bar{s}s\rangle$   quark-antiquark condensates in hadron phase. The contribution to the pressure due to a particle of mass $m_h$, baryon charge $B$, isospin $I_3$, strangeness $S$,  and degeneracy $g$ is given by
\begin{eqnarray}
\Delta p=\frac{g\, m_h^2\, T^2}{2\pi^2} \, \sum_{n=1}^\infty \,\frac{(-\eta)^{n+1}}{n^2} \,\exp\left(n\frac{B\mu_B - I_3\mu_I - S \mu_S}{T} \right) \, K_2\left(n\frac{m_h}{T} \right),   \label{p1}
\end{eqnarray}
where $K_n(x)$ is the modified Bessel function. In hadrons, the isospin is an almost exact symmetry.  

The quark-antiquark condensates are given by the derivative of Eq. (\ref{p1}) with respect to the constituent quark masses
\begin{eqnarray}
\langle\bar{q}q\rangle &=&\langle\bar{q}q\rangle_0+ \sum_h \frac{\partial   m_h}{\partial m_q} 
\frac{\partial \Delta p}{\partial m_h}, \nonumber \\ 
\langle\bar{s}s\rangle &=&\langle\bar{s}s\rangle_0+ \sum_h  \frac{\partial m_h}{\partial m_s} \frac{\partial \Delta
  p}{\partial m_h},  \label{qqHRG}
\end{eqnarray}
where $\langle\bar{q}q\rangle_0$ and $\langle\bar{s}s\rangle_0$ are light and strange quark-antiquark condensates in  vacuum, respectively \cite{Tawfik:2005qh}.  It was found that at small chemical potential the strange quark-antiquark condensate is larger that the light one. At large chemical potential, such difference gradually diminishes. 

Some authors still prefer to take into account repulsive ({\it electromagnetic}) van der Waals interactions in order to compensate the strong interactions in hadron matter \cite{Tawfik:2013eua}. Accordingly, each resonance constituent is allowed to have an {\it eigen}-volume. Thus, such total volume should be subtracted from the fireball volume or that of the heat bath. Also, considerable modifications in thermodynamics of hadron gas including energy, entropy and number densities are likely. The hard-core radius of hadron nuclei can be related to the multiplicity fluctuations. 

About ten years ago, Tawfik derived $S$-matrix for the HRG model \cite{Tawfik:2004sw}, which describes the scattering processes in the thermodynamical system~\cite{Dashen:1969mb}. Accordingly, Eq.~(\ref{eq:lnz1}) can be written as an expansion of the fugacity term   
\begin{eqnarray}
\ln\, Z^{(int)}(V,T,\mu) &=& \ln\, Z^{(id)}(V,T,\mu) + \sum_{\nu=2}^{\infty} a_{\nu}(T) \exp(\mu_{\nu}/T). \label{eq:p2}
\end{eqnarray}
where $a_{\nu}(T)$ are the virial coefficients and the subscript $\nu$ refers to the order of multiple-particle interactions.
\begin{eqnarray}
a_{\nu}(T) &=& \frac{g_r}{2\pi^3} \int_{M_{\nu}}^{\infty}dw\; \exp\{-\varepsilon_r(w)/T\}\;
\sum_l(2l+1)\frac{\partial}{\partial w}\delta_l(w). \label{eq:p3}
\end{eqnarray}
The sum runs over the spatial waves. The phase shift $\delta_l(w)$ of two-body inelastic interactions, for instance, depends on the resonance half-width $\Gamma_r$, spin and mass of produced resonances,  
\begin{eqnarray}
\ln\, Z^{(int)}(V,T,\mu) &=& \ln\, Z^{(id)}(V,T,\mu) + \frac{g_r}{2\pi^3}\int_{M_{\nu}}^{\infty}dw  \frac{\Gamma_r\; \exp\{(-\varepsilon_r(w)+\mu_r)/T\}}{(M_r-w)^2+\left(\frac{\Gamma_r}{2}\right)^2}. \label{eq:p4}  
\end{eqnarray}
In Eq.~(\ref{eq:p4}), by replacing $\mu$ by $-\mu$, the anti-particles are taken into consideration. For a narrow width and/or at low temperature, the virial term decreases so that the {\it non-relativistic} ideal partition function of hadron resonances with effective masses $M_{\nu}$ is obtained. This means that, the resonance contributions to the partition function are the same as that of massive {\it free} resonances. At temperatures comparable to $\Gamma_r$, the effective mass approaches the physical one. Thus, we conclude that at high temperatures, the strong interactions are taken into consideration via heavy resonances, Eq.~(\ref{eq:lnz1}), i.e. Hagedorn picture. We therefore utilise the grand canonical partition function, Eq.~(\ref{eq:p2}), without any corrections.

In order to verify this picture, Tawfik checked the ability of HRG with finite-volumed constituents in reproducing lattice QCD thermodynamics \cite{Tawfik:2013eua}. At radius $r>0.2~$fm, the disagreement becomes obvious and increases with increasing $r$. At high temperatures, the resulting thermodynamics becomes {\it non}-physical. It was concluded that the excluded volume seems to be practically irrelevant. It has e negligible effect, at $r\leq 0.2~$fm. On the other hand, a remarkable deviation from the lattice QCD calculations appears, especially when the radius $r$ become large. 

In the present work, the chiral parameters, $M(T)$ and $\Delta_ {q,s}(T)$, see section \ref{sec:Results}, are extracted from HRG assuming fully and partially chemical non-equilibrium \cite{Tawfik:2014dha}. There is no difference, when $\gamma_S=1.0$ and when it is allowed to have values different than unity, where $\gamma_q$ and $\gamma_S$ refer to non-equilibrium treatment or occupation factors for light and strange quarks, respectively. These two parameters enter Eq. (\ref{eq:lnz1}) after raising them to exponents reflecting the light and strange quarks contents of $i$-th hadron. They are identical to the fugacity factor and therefore are multiplied to the exponential function.


\section{Results}
\label{sec:Results}

A systematic comparison between PLSM, PNJL, QLSM and HRG is presented. It intends to calculate two chiral quantities, the order parameter $M(T)$ and the normalized net strange non-strange condensate $\Delta_{q,s}(T)$. The results shall be confronted to the lattice QCD simulations \citep{LQCD1,Schmidt:2010ss,Borsanyi:2010zi}. The comparison with the lattice should signal which model is close to the lattice and on other hand offers differentiation between the SU(3) effective models, themselves.

\subsection{Chiral order-parameter $M(T)$}
\label{subsec:Results1}

\begin{figure}[htb]
\includegraphics[width=8.cm,angle=-90]{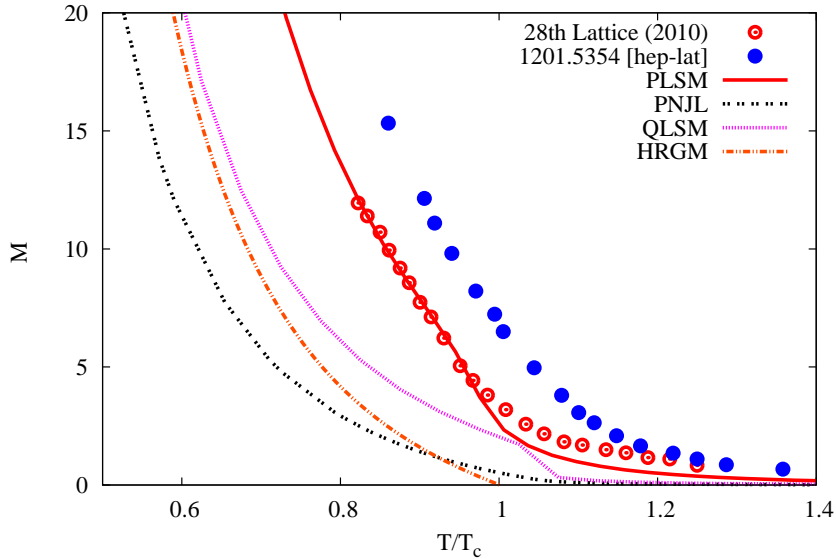}
\label{sec:QCDcomp}
\caption{The thermal behaviour of the dimensionless chiral order-parameter, $M$, calculated  as function of temperature from the four SU(3) effective models, PLSM (solid curve), QLSM (dotted curve), PNJL (double-dotted curve) and HRG (dash-double-dotted curve) and compared with the lattice QCD calculations (solid circles) \cite{LQCD1} and (open circles) \cite{Schmidt:2010ss} at $m_l/m_s=0.037$. 
\label{fig:M} 
}
\end{figure}

The chiral order-parameter $M(T)$ was originated in lattice techniques \cite{Mqcd}. The latter calculates dimensionless quantities in units of lattice spacing rather than physical units. The lattice spacing can then be converted into the physical  units. $M(T)$ relates the light quark condensate to the strange quark mass
\begin{eqnarray} 
M(T) &=& m_s\, \dfrac{\langle\bar{\psi}\psi\rangle_{l}}{T^{4}}, \label{Eq:M}
\end{eqnarray}
where $m_s$ is the strange quark physical mass which fixed here to $138~MeV$ in order to get the ratio of light and strange quark masses $m_l/m_s=0.037$.  Also, we  notice that the dimensionless $M(T)$ depends on the thermal behavior of the light quark condensate $\langle\bar{\psi}\, \psi_{l}\rangle$. In  lattice QCD, the chiral condensate  remains finite. But it contains contributions which would diverge in the continuum limit. Therefore, it requires renormalization, in particular an additive and multiplicative renormalization. In order to remove - at least - the multiplicative renomalization factor, we take into consideration Eq. (\ref{Eq:M}) as a definition for the order parameter. The light quark condensate itself can be calculated from the potential,  PLSM: Eq. (\ref{plsm:omega}), 
QLSM: Eq.  (\ref{Pq}),
PNJL: Eq. (\ref{PNJLpotuntial}) and
HRG: Eq. (\ref{qqHRG}).
Accordingly, we estimate Eq. (\ref{Eq:M}) from the four models and then compare them with the lattice QCD calculations, Fig. \ref{fig:M}. We find that this chiral order-parameter in the SU(3) effective models and first-principle lattice QCD simulations \cite{LQCD1,Schmidt:2010ss} rapidly decreases with increasing $T$. 

Comparing with the lattice QCD \cite{LQCD1}, the best agreement is found with PLSM, but PLSM underestimates the recent lattice QCD \cite{Schmidt:2010ss}. In fact, the lattice calculations \cite{Schmidt:2010ss} lay on top of all curved from the SU(3) effective models. This might be originated to the specific configurations of the lattices and the actions implemented in each simulations, section \ref{sec:QCDcomp}. The other effective models lay below the two sets of lattice calculations.

The four models PLSM, QLSM, PNJL, HRG and different sets of lattice calculations have different critical temperatures. In Tab. \ref{tab:1}, we list out the critical temperatures corresponding to each order parameter. In determining the pseudocritical temperatures, $T_{\chi}$, different criteria are implemented. They are not only quite unorthodox but they are distinguishable from each other. Further details shall be elaborated in Section \ref{subsec:Results2}.

\begin{table}
\begin{center}
\begin{tabular}{||c||c|c||}
\hline\hline
  & $T_{\chi}$ [MeV]  & Order Parameter \\ \hline\hline
PLSM &  $164$  &  crossing of ($\sigma_x$, $\sigma_y$) and  ($\phi$, $\bar{\phi}$) \\ \hline
QLSM & $200$ & ($\sigma_x$, $\sigma_y$)  and largest fluctuation in $m_2/\mu^2$  \\ \hline
PNJL &  $217$ & crossing of ($\sigma_x$, $\sigma_y$) and ($\phi$, $\bar{\phi}$) \\ \hline
HRG &   $184$ & vanishing $\langle \bar{\psi}\psi\rangle$-condensate \\ \hline
LQCD \cite{LQCD1} & $156$ at $m_s/m_l=0.037$ & sudden drop in $M(T)$ and $\Delta_{l,s}(T)$ \\ \hline 
LQCD \cite{Schmidt:2010ss} &  $165-170$ & sudden drop in $M(T)$ and $\Delta_{l,s}(T)$ \\ \hline
LQCD \cite{Borsanyi:2010zi} &  $165$ & sudden drop in $M(T)$ and $\Delta_{l,s}(T)$ \\ \hline
\end{tabular}
\caption{The pseudocritical temperatures $T_{\chi}$ as calculated from PLSM, QLSM, PNJL, HRG and the different sets of lattice QCD  calculations.  \label{tab:1}  }
\end{center}
\end{table}

\subsubsection{A short comparison between the two sets of lattice QCD calculations}
\label{sec:compLQCD}

Refs. \cite{LQCD1,Schmidt:2010ss} presented results for $2+1$ quark flavors, where all systematics are controlled, the quark masses are set to their physical values and the continuum extrapolation is carried out. Larger lattices and a Symanzik improved gauge besides a stout-link improved staggered fermion action are implemented.  Depending on the exact definition of the observables, the remnant of the chiral transition is obtained at $T_c=150$~MeV. Extending these results, the transition temperature  was also determined for small non-vanishing baryonic chemical potentials. At high temperatures, the lattice pressure is found $\sim 30\%$ lower than the Stefan-Boltzmann limit.

Ref. \cite{Borsanyi:2010zi} used $2+1$ quark flavors with physical strange quark mass and almost physical light quark masses. The calculations have been performed with two different improved staggered fermion actions, the asqtad and p4 actions.  Overall, a good agreement between results obtained with these  two $O(a^2)$ improved staggered fermion discretization schemes is found. At high temperatures, the lattice pressure is $\sim 14\%$ lower than the Stefan-Boltzmann limit.

From this short comparison, we find that:
\begin{itemize}
\item \cite{LQCD1,Schmidt:2010ss} implement Symanzik improved gauge and stout-link improved staggered fermion action. The resulting pressure is found $\sim 30\%$ lower than the Stefan-Boltzmann limit.
\item \cite{Borsanyi:2010zi} uses improved staggered fermion actions; the asqtad and p4 actions. The resulting pressure is $\sim 14\%$ lower than the Stefan-Boltzmann limit.
\end{itemize}

\subsubsection{Couplings in PLSM}
In the effective models, the parameters, especially the couplings, are very crucial for the outcome of the calculations. One of the motivations for the present work is the failure of PLSM \cite{Tawfik:2014uka} in reproducing the lattice QCD results \cite{LQCD1,Schmidt:2010ss,Borsanyi:2010zi} even with large coupling $g$. In Ref. \cite{Tawfik:2014uka}, $g$ ranges between $6.5$ and $10.5$. The first value was enough to reproduce the lattice QCD calculations, PRD80, 014504 (2009) and PLB730, 99 (2014). Increasing $g$ to $10.5$ does not enable PLSM to reproduce the other lattice simulations \cite{LQCD1,Schmidt:2010ss,Borsanyi:2010zi}. Furthermore, through fitting with lattice QCD calculations and experiments, the parameters of PLSM can be estimated. This was described in details in Ref.  \cite{Tawfik:2014uka,Schaefer:2008hk}. 

Scope of the present script is the regeneration for the lattice QCD calculations \cite{LQCD1,Schmidt:2010ss,Borsanyi:2010zi}.  In the present work, we tackle this problem through comparison with various effective models. In doing this, we have modified LSM and present systematic analysis for two order parameters. We have added to LSM the gluonic sector of the quasi-particle model. This is the essential original proposal of the present script. Thus, waiving details about PLSM itself is though as legitimated. But for a complete list of the PLSM parameters, the readers are kindly advised to consult  \cite{Tawfik:2014uka,Schaefer:2008hk}.

\subsection{Normalized net-strange condensate $\Delta_{q,s}(T)$}

\begin{figure}[htb]
\includegraphics[width=8cm,angle=-90]{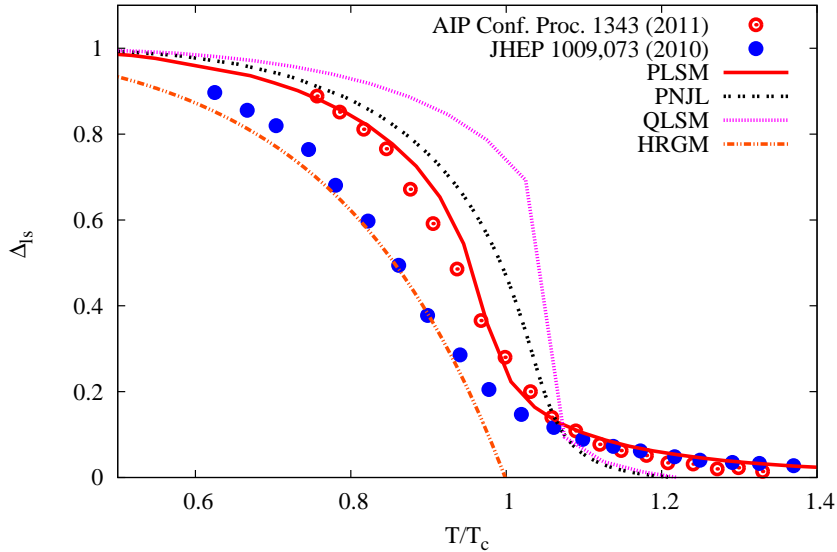}
\caption{The thermal dependence of $\Delta_{q,s}$ calculated from PLSM (solid curve), QLSM (dotted curve), PNJL (double-dotted curve) and HRG (dash-double-dotted curve) and compared with the lattice QCD calculations (solid circles) \cite{LQCD1} and  (open circles) \citep{Borsanyi:2010zi} at $m_l/m_s=0.037$. \label{fig:Delta} 
}
\end{figure}

Another dimensionless quantity shows the difference between non-strange  and strange condensates. 
\begin{eqnarray}
 \Delta_ {q,s}(T) &=& \dfrac{\langle\bar{q}q\rangle - \dfrac{m_q}{m_s} \langle\bar{s}s\rangle}{\langle\bar{q}q\rangle_{0} - \dfrac{m_q}{m_s} \langle\bar{s}s\rangle_{0}},\label{Eq:Delta}
\end{eqnarray}
where $\langle\bar{q}q\rangle$ ($\langle\bar{s}s\rangle$) are non-strange (strange) condensates, and $m_q$ ($m_s$) are non-strange (strange) masses. Using Ward identities and Gell-Mann-Oakes-Renner relation, expression (\ref{Eq:Delta}) might be given in terms of pion and kaon masses and their decay constants \cite{GMOR}. Accordingly, the final results might be scaled but their thermal behavior remains unchanged.
The lattice QCD calculations for $\Delta_{q,s}(T)$  (solid circles) \cite{LQCD1} and  (open circles) \citep{Borsanyi:2010zi} are compared with the calculations from PLSM (solid curve), QLSM (dotted curve), PNJL (double-dotted curve) and HRG (dash-double-dotted curve) in Fig. \ref{fig:Delta}. 

It is obvious that PLSM agrees with the lattice results \cite{Borsanyi:2010zi} at low and also at high temperatures. Its ability to reproduce the other set of lattice results \cite{Schmidt:2010ss} is limited to the high temperatures. This might be originated in the difference between the two sets of lattice QCD simulations, section \ref{sec:compLQCD}. The HRG model agrees well with this lattice calculations \cite{Schmidt:2010ss}. It is apparent that such agreement is limited to temperatures below the critical value due to the limited applicability of the HRG model. The remaining two models PNJL and QLSM show qualitative thermal behavior, as that from the other effective models and lattice calculations, which can be described by large plateau at low temperatures, around the critical temperature the values of  $\Delta_{q,s}(T)$ decrease rapidly and at high temperature, $\Delta_{q,s}(T)$ vanishes but very slowly. Both models are closer to \cite{Borsanyi:2010zi} rather than to \cite{Schmidt:2010ss}

Both Figs. \ref{fig:M}  and \ref{fig:Delta} show the PNJL model and HRG model describe much better the LQCD data for the magnetization and normalized net strange condensate, respectively, than for the chiral condensate. One should bear in mind that the magnetizations have been simulated in a different lattice that the one for the net strange condensate. Unfortunately, both quantities are not available from the same lattice simulation.

\subsection{QCD chiral phase-diagram}
\label{subsec:Results2}

For mapping out the QCD chiral phase-diagram, various approaches are available. From PLSM and PNJL, as they possess two order parameters; one for strange and one for non-strange chiral condensates, hints about QCD chiral phase transition can be analysed. Furthermore, PLSM and PNJL possess deconfinement order parameter because of the Polyakov loop potential. Therefore, from strange and non-strange chiral condensates, a dimensionless quantity reflecting the difference between both condensates, $\Delta_{q,s}(T)$, can be deduced as function of temperature at fixed baryon chemical potential. Apparently, this signals the QCD chiral phase-transition. At the same value of baryon chemical potential, we can also deduce the deconfinement order-parameter as function of temperature. At a fixed baryon chemical potential, the thermal dependence of these two quantities intersect with each other at a characterizing point representing the phase transition. When repeating this procedure at different values of the baryon chemical potentials, we get a set of points representing the QCD phase-diagram. The results are given in Fig. \ref{fig:C-PD}, as solid curve for PLSM and dotted curve for PNJL. 

For the QCD chiral phase-diagram from QLSM, we implement another method. As no Polyakov loop potential is included, the QCD chiral phase-diagram is characterized by the higher-order moments of particle multiplicity \cite{Tawfik:2014bna}, which are assumed to highlight various types of fluctuations in $T$ and $\mu$. Therefore, we utilize the possible fluctuations accompanying normalized second-order moment \cite{Tawfik:2014bna} in mapping out the QCD chiral phase-transition. The problematic of determining pseudocritical temperature from the second moment has been discussed in Ref. \cite{Karsch2009a}. Accordingly, we observe that the peaks corresponding to different temperatures are conjectured to be characterized by different values of the baryon chemical potentials, where the QCD chiral phase-transition is conjectured to occur. We analyse this dependence at different values of the temperature $T$. Then, we follow the scheme to determine $T$ and $\mu$, which is characterized by maximum $m_2/\mu^2$, where $m_2$ is the second-order moment of  the particle multiplicity. The results are  illustrated in Fig. \ref{fig:C-PD}, as dash-dotted curve.

For the  HRG model, we map out the QCD chiral phase-diagram by utilizing the quark-antiquark condensate as order parameter \cite{Tawfik:2005qh}. It is assumed that the thermal dependence of the quark-antiquark condensate remains finite in the hadronic phase, but vanishes at temperature higher than the critical chiral-temperature. The results are given in Fig. \ref{fig:C-PD}, as well, as double-dotted curve.

We can now shortly summarize the methods implemented to determine the pseuodcritical temperatures: 
\begin{itemize}
\item PLSM and PNJL: due to chiral and deconfinement phase transitions for light and strange quarks, $\Delta_{q,s}(T)$ is determined as function of temperature at a fixed baryon chemical potential, $\mu$. This signals the QCD chiral phase-transition. At the same value of $\mu$, the deconfinement order-parameter can be studied as function of temperature, as well. Then, the thermal  dependence of these two quantities is conjectured to intersect with each other at a characterizing point. When repeating this procedure for different values of $\mu$, a set of points of pseudocritical temperatures $T_{\chi}$ and   $\mu$ can be deduced.
\item QLSM: the normalized second-order moment of particle multiplicity is implemented in mapping out the QCD chiral phase-transition. Peaks corresponding to different temperatures are conjectured to be characterized by  different values of  $T_{\chi}$ and  $\mu$.
\item HRG: the quark-antiquark condensates are implemented as order parameters. At vanishing and finite $\mu$, the thermal dependence of the quark-antiquark condensate remains finite in the hadronic phase and vanishes at temperature higher than the critical chiral-temperature,  $T_{\chi}$. 
\end{itemize}

We observe that the chiral boundary from PLSM (solid curve) is positioned within the upper band of the lattice QCD calculations  \cite{KarschA,LQCDA} and agrees well with the freeze-out results deduced from the STAR BES measurements (symbols) \cite{Tawfik:2013bza}. The temperatures calculated from the HRG model by using the quark-antiquark condensate as the order parameter (double-dotted curve) \cite{Tawfik:2005qh} is higher than the chiral temperatures from the PLSM and the freeze-out temperatures calculated in the lattice QCD (band) and from the STAR BES measurements (symbols). Despite this difference, the corresponding $T$-$\mu$ sets are very similar to that of the PLSM. The results from PNJL and QLSM are higher than that from the HRG model.

\begin{figure}[htb]
\includegraphics[width=8cm,angle=-90]{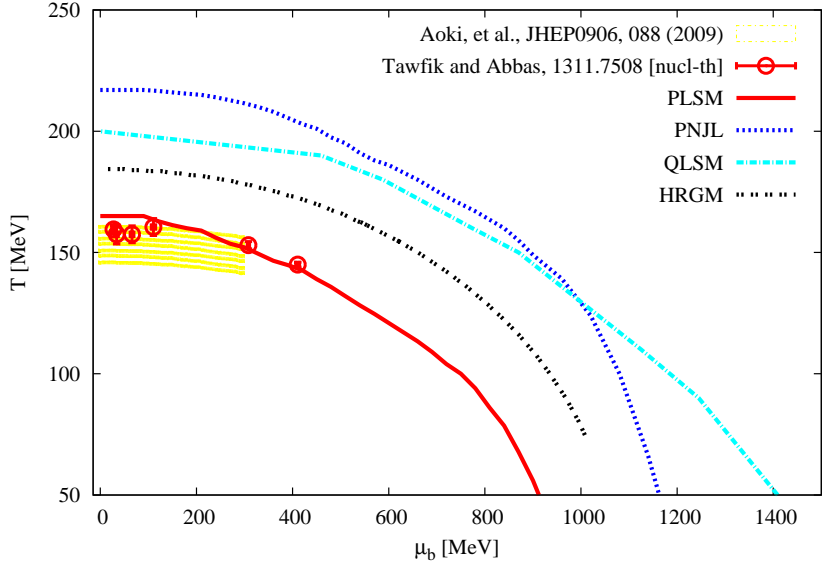}
\caption{The PLSM $T$-$\mu$ chiral phase-diagram (lines with points), with which the freeze-out parameters deduced from lattice the QCD calculations \cite{KarschA,LQCDA} (band) and that from different the thermal models \cite{Tawfik:2013bza,SHMUrQM} (symbols) are compared. 
\label{fig:C-PD}}
\end{figure}

\section{Conclusions and outlook}
\label{sec:conclusion}

In the present work, we report on a systematic comparison between PLSM, PNJL, QLSM and HRG in generating the chiral quantities, order parameter $M(T)$ and normalized net strange and non-strange condensates $\Delta_{q,s}(T)$. Furthermore, we confront the results deduced from the four effective models to the recent lattice QCD calculations in order to distinguish  between the models and to interpret the first-principle lattice QCD calculations.

For the order parameter $M(T)$, the best agreement is found with PLSM, while the recent lattice QCD \cite{Schmidt:2010ss} lay on top of all curves. This might be understood from the lattice configurations and the actions implemented in the simulations. The other effective models lay below the two sets of the lattice calculations. We notice that the effective PLSM, QLSM, PNJL and HRG and the different sets of the lattice calculations have different critical temperatures, Tab. \ref{tab:1}.

For the normalized net strange and non-strange condensates $\Delta_{q,s}(T)$, PLSM again gives an excellent agreement with the lattice results \cite{Borsanyi:2010zi} at low and high temperatures. But its ability to reproduce the lattice simulations \cite{Schmidt:2010ss} is limited to high temperatures. This might be originated in the difference between the two sets of lattice QCD simulations.  Furthermore, we find that the HRG model agrees well with this the lattice QCD calculations \cite{Schmidt:2010ss}. It is apparent that this is  restricted to temperatures below the critical value. The effective models PNJL and QLSM show the same qualitative thermal behavior. There is a large plateau at low temperatures. Around the critical temperature the values of  $\Delta_{q,s}(T)$ decrease, rapidly. At high temperature, $\Delta_{q,s}(T)$ vanishes but very slowly. The effective models PNJL and QLSM are closer to \cite{Borsanyi:2010zi} rather than to \cite{Schmidt:2010ss}.

In light of this, we conclude that the PLSM reproduces $M(T)$ and $\Delta_{q,s}(T)$, well. The HRG model is able to reproduce $\Delta_{q,s}(T)$, while PNJL and QLSM seem to fail. These features and differences are present in the chiral phase-diagram, Fig. \ref{fig:C-PD},  as well.

In section \ref{subsec:Results2}, we have introduced the various order parameters used in the different models in order to deduce $T$ and $\mu$ of the  QCD chiral phase-transition. The strange and non-strange chiral condensates and  the Polyakov loop potentials are utilized in PLSM and PNJL. The thermal dependence of these two quantities are assumed to intersect with each other at a characterizing point representing the QCD chiral phase-transition. For QLSM,  no Polyakov loop potential is included in, therefore, the chiral phase-diagram is characterized by the higher-order moments of the particle multiplicity. The possible fluctuations accompanying the normalized second-order moment are assumed to map out the QCD chiral phase-transition. For the HRG model, we utilize the quark-antiquark condensates as order parameter. 

Again, we find that the PLSM chiral boundary (solid curve) is located within the upper band of the lattice QCD calculations and agrees well with the freeze-out results deduced from the experiments and the thermal models (symbols). It is obvious that the chiral temperature calculated from the HRG model is larger than that from the PLSM. This is also larger  than the freeze-out temperatures calculated in the lattice QCD (band) and from the experiments and the thermal models (symbols). Despite this difference, the corresponding $T$ and $\mu$ sets are very similar to that from the PLSM. This might be explained as follows. The $T$ and $\mu$ are calculated using different order parameters; in HRG vanishing quark-antiquark condensate but in the PLSM crossing (equalling) the chiral condensates and the Polyakov loop potential. The latter  assumes that the two phase transitions;  the chiral and  the deconfinement, occur at the same temperature. The earlier deals with the chiral phase-transition independent on the  confinement-deconfinement one.

The results from the two model PNJL and QLSM show the same qualitative behavior. The chiral temperatures are higher than that from the PLSM and HRG. This might be interpreted due the heavy quark masses implemented in both models.

Any model comparison with lattice results should span as much as possible of the parameter space. Even with the narrow parameter space explored in the present paper, we would like to highlight that the results are limited this. But, with reference to previous work \cite{Tawfik:2014uka}, the parameters alone are not able to explain the diversity with the results in this study. We have to attack essential components of LSM and integrate gluonic sector taken from the quasi-particle models.

\section*{Acknowledgement}
The present work was supported by the World Laboratory for Cosmology And Particle Physics (WLCAPP) http://wlcapp.net/. 
The authors are very grateful to the anonymous referee for his/her very constructive comments, suggestions and even criticisms, which helped a lot in improving the manuscript!



\end{itemize}

\end{document}